\documentclass[12pt,preprint]{aastex}
\begin{document}
\title {DIRECT MICROLENSING-REVERBERATION OBSERVATIONS OF THE INTRINSIC 
MAGNETIC STRUCTURE OF AGN IN DIFFERENT SPECTRAL STATES - A 
TALE OF TWO QUASARS}
\author{ Rudolph E. Schild\footnote{Center for Astrophysics, 60 Garden
Street,
Cambridge, MA 02138}  , Darryl J. Leiter\footnote{Marwood Astrophysical
Research Center, Charlottesville, VA 22963}  , and Stanley L. 
Robertson\footnote{
Dept. of Physics, Southwestern Oklahoma State University,
Weatherford, OK 73096}}

\begin{abstract}
We show how direct microlensing-reverberation analysis performed on two
well-known Quasars (Q2237 - The Einstein Cross and Q0957 - The Twin) can
be used to observe the inner structure of two quasars which are in
significantly different spectral states. These observations allow us to
measure the detailed internal structure of quasar Q2237 in a radio quiet
high-soft state, and compare it to quasar Q0957 in a radio loud low-hard
state. We find that the observed differences in the spectral states of
these
two quasars can be understood as being due to the location of the inner
radii of their accretion disks relative to the co-rotation radii of
rotating
intrinsically magnetic supermassive compact objects in the centers of
these quasars.

\end{abstract}
\section{Introduction}
An important fundamental issue confronting the study of AGN and Galactic
Black Hole Candidates (GBHC) is the fact that these objects are observed
to exist in different spectral states. The phenomenon of spectral state
switching was first discovered in neutron star X-ray binaries where it was
explained in terms of magnetic propeller effects associated with the
interaction of the intrinsic magnetic moment of the of the neutron star
with its accretion disk (Ilarianov \& Sunyaev, 1975) 
Spectral state switching was later observed in GBHC where it was found
(Robertson and Leiter, 2002, 2003, 2004, 2005) that it could be explained 
in terms of magnetic propeller effects
associated with the interaction of the intrinsic magnetic moment of the
GBHC, generated by the central highly red shifted "Magnetospheric Eternally
Collapsing Object" (MECO) within it, with the GBHC accretion disk. In these
papers it was shown that a unified explanation of spectral properties of
GBHC in Radio Loud Low Hard States and Radio Quiet High Soft States can be
given as being dynamically generated 
by the relative location of the inner
magnetospheric radii of their accretion disks, with respect to the
co-rotation radius associated with the intrinsically magnetic MECO
at their centers. A qualitative description of this switching has also been
given by Bellon1 (2007).

Recent observations, associated with the effects of intrinsic magnetic
propeller interactions generated by MECO in quasars, have been discussed by
Schild, Leiter, and Robertson, 2006, hereafter referred to as SLR06, in
which intensive gravitational microlensing observations of quasar Q0957
were described which permitted a reconstruction of this quasars inner
radiation emitting structures. Surprisingly these observations did not
reveal the expected accretion disk extending in close to the central
object.
Instead the inner accretion disk was observed to be truncated at a very
large inner radius. The hot inner edge of the truncated accretion disc
appears to be dominated by a very thin plasma annulus, whose thermal
radiation is responsible for the ``small blue bump'' of the quasar spectral
energy distribution. In addition, large conical outflow
structures were observed about ten times further out from the disk's
inner radius.

The overall morphology of these dynamic structures was found to be similar
to features revealed in simulated accretion flows onto a compact Young
Stellar
Object which contains a central rotating magnetic dipole object inside of
the accretion disk. Since black holes do not contain observable intrinsic
dipole fields, such dynamic structures cannot be generated by black
hole-accretion disk models. Hence the conclusion was that we were observing
the physical effects of an intrinsic magnetic propeller interaction of a
central compact magnetic object in Q0957, and that instead of a black hole
the super-massive compact object in this quasar was a MECO. (For a review
of the observational and theoretical arguments involved in reaching this
conclusion about quasar Q0957 see Appendix I).

Current observations of galactic black hole candidates (GBHC) and
active galactic nuclei (AGN) have begun to suggest that the existence of
different spectral states, associated with the presence or absence of
intrinsic magnetic propeller effects, may play an important role in
explaining the broad range in observed AGN properties (radio and hard
X-ray
emission, UV optical brightness). In this context we will show in this
paper
how the application of combined  direct  microlensing-reverberation
analysis  on two well known multiple image based quasars (Q2237 - The
Einstein  Cross and Q0957 -The Twin) can be used to determine the detailed
inner structure of two quasars which are in significantly  different
spectral states.

Sections 1-3 of this paper will show how this combined observational
technique allows us to measure the internal structure of quasar Q2237 in a
radio quiet high-soft state, and compare it to the internal structure
which has been observed in quasar Q0957 in a radio loud 
low-hard state (SLR06).
In sections 4 and 5 we then proceed to analyze the data associated with
these observed internal quasar structures of these two quasars in the
context of mass scaled versions of recently developed theoretical models
(Robertson and Leiter 2002, 2003, 2004, 2005) of 
Galactic Black Hole Candidates
(GBHC) that have the ability to describe spectral state switching between 
a "Radio Loud Low-Hard State" (low level of soft X-ray emission and strong
hard X-ray emission plus strong radio emission) and a "Radio Quiet
High-Soft
State" (high level of soft X-ray emission but low level of hard X-ray
emission, with only a weak level radio emission). In this manner we show
how direct microlensing-reverberation analysis allows us to obtain a deeper
insight into the fundamental physics which underlies the
observational fact that quasars, AGN, and GBHC are observed to be able to
occupy a common set of different spectral states. These spectral states,
which have properties that scale with the mass of the central compact
object, can be classified by observable characteristics such as: strong or
weak radio emission, hard or soft X-ray fluxes, and the presence or
absence of broad relativistic $Fe-K_(\alpha)$ lines.

For GBHC at low red shift the internal structures responsible for the
different spectral states are too small to be resolved at the wavelengths
available with current technology. On the other hand for quasars, which are
observed at cosmological distances, it has been found that a combination of
direct reverberation, microlensing, and nano-lensing techniques are able to
resolve the internal quasar structures associated with their spectral states
(SLR06).

While individual GBHC objects are observed to switch between the spectral
states on time scales of minutes, for AGN and quasars the time scale for
the
spectral switching process is expected to be the order of years; hence the
individually observed AGN will appear to be "frozen" in one state or the 
other.

Reverberation analysis has proved to be an important technique for the
study
of Active Galactic Nuclei (AGN) because it can define the characteristics of
the complex internal structures in the highly relativistic inner regions
which are still too compact to be resolved by direct imaging. The original
application by Peterson et al (1991, 1993) showed that a lag between
brightness changes of the continuum and emission line regions can be
interpreted to indicate the length scale between the regions.
Reverberation
of the UV-optical continuum light with itself was also discovered, and
shown
in Q0957 to indicate the existence of internal structure on time scales of
50 days (Schild \& Thomson 1997; S05). In addition, 
reverberation of the $2{\micron}$ emission against
the continuum in NGC 5548 showed that the near-IR emission comes from a
region just beyond the broad line region (Suganuma et al, 2004; see similar
result for NGC4151 in Swain et al, 2003; Minezaki et al 2004).

The reverberation analysis in Q0957 seems particularly interesting because
it shows the existence of several non-standard forms of internal
structures
at lags that seem consistent with emission from the Elvis structures that
give rise to the broad emission lines (SLR06).
The existence of the pattern stream was inferred from auto-correlation
analysis 
applied to the long brightness record accumulated by Schild and colleagues
(Schild \& Thomson, 1995) for the gravitationally lensed multiple image 
Q0957+561 A,B quasar. A
combination of auto-correlation analysis and reverberation methods was
required in the analysis of Q0957 since its pattern of intrinsic
brightness
fluctuations was quite complex, with several pattern streams overlapping
to an extent that prevented single patterns from being easily recognized.

Fortunately the growing brightness record accumulated from photometric
monitoring of the Einstein Cross 4-image gravitational lens Q2237+0305 has
shown an unambiguous pattern of brightness peaks that confirm the
inferences
previously made for Q0957, and allows the relative brightness of the peaks
to be estimated. This allowed us to clarify the relationship of the peaks
to the internal quasar structures, and also provides unique information on
the
orientation of the quasar in space with respect to the plane of the sky. In
the following section, we will show that in the Q2237+0305 quasar lens, an
initial outburst in the central region is followed by reverberations
interpreted as fluorescent repeats of the central disturbance, since within
reasonable limits they
have the same amplitude and time scales. From
these we will estimate the size, spacing, and orientation of the
UV-optically  luminous  structures in the Q2237 quasar. We will then show
that our size estimates are consistent with direct microlensing estimates of
quasar size. In particular our reverberation size scale estimates will be
shown to be consistent with the Vakulik et al (2007) estimates of the
brightness and size scales of the Q2237 inner UV-optical structure.

\section{Determination of Quasar Intrinsic Brightness Peak Lags}

Our approach to the determination of the Q2237 structure parameters follows
closely the procedure outlined for the Q0957 quasar (S05; SLR06). 
The procedure is a variant of the
reverberation analysis first applied by Peterson et al (1991,1993) to NGC
5548; however their observations and analysis were of the reverberation
between the optical continuum and the emission lines. In our implementation
for Q0957, reverberation of the continuum emission by itself was discovered
by Schild and Thomson (1997), wherein
it was noticed that multiple peaks are found in the autocorrelation
estimates of the long Q0957 brightness record compiled by Schild and
Thomson (1995). 
The actual succession of peaks evidencing the reverberations have
not been identified in the Schild et al brightness data records, because of
irregular data sampling, and because many overlapping patterns are seen
throughout the Q0957 brightness record. However in Q2237, where 4 quasar
images and the central galaxy are clearly seen in the well-known Einstein
Cross pattern, identical brightness peaks are clearly evident, and these
have been exploited by Vakulik et al (2006) to determine the very short
(approximately 1 day) time delay lags of the overall strong gravitational 
lensing configuration. The excellent photometric data and their
plot can be seen at (http://www.astrouw.edu.pl/~ogle/ogle3/huchra.html)
(Wozniak et al. (2000).
In their colored plot of the image brightness, the first brightening pulse
must have occurred at approximately HJD-2450000 = 2550, although the
actual
mid-point and profile of the initial event are obscured because of poor
sampling at the time. The event is signaled by a brightness increase of
0.3 magnitudes for a duration of approximately 150 days. This was
especially well seen in the data for image A (green data points).
The A image is apparently undergoing a sustained period of low amplitude
microlensing whose peak was previously seen in the OGLE I data set
(http://bulge.princeton.edu/~ogle/ogle2/huchra.html)(Wozniak et al 2000).
Although the peak is defined by brightness measured near the beginning and
end only in the OGLE III data set, additional data from Koptelova et al
(2004) show that the brightness was constant during that observing season
for all four quasar images.

The new quasar intrinsic brightening event can probably also be
recognized in the B, C, and D images, which causes us to conclude that it
is an intrinsic quasar brightening event, not a microlensing artifact.
The first peak was followed 415 days later by another peak of comparable
brightness at 2965 days, again particularly well seen in the A image green
brightness record. This nearly simultaneous rise in all four images was
the
observational basis for the determination of the time delays of the four
images by Vakulik et al, 2006. These authors analyzed the V-band OGLE III
data but also their own R-band data, which had somewhat lower brightness
 peak amplitudes. The lags determined were approximately one day, as
expected. The brightness amplitude, about 0.3 magnitudes, and event
duration, approximately 100 days, were all comparable to, but slightly
lower than, the initial peak. A third peak can barely be recognized at
approximately HJD - 2450000 = 3300, but at only half the amplitude of the
first two peaks. Nevertheless the duration of the event is the same as for
the first two peaks, although the third peak cannot be recognized at all in
image B because a microlensing event has evidently distorted the brightness
curve. A fourth and somewhat brighter peak can probably be recognized at
HJD - 2450000 = 3700 days, but microlensing of both the A and B images 
combine with poor sampling because  of seasonal observing effects, to
observe this feature. In Q2237, a fifth and final pulse might be
recognizable, but it appears to be too weak to distinguish it from the
noise associated with microlensing confusion and seasonal data effects.
It is evidently weak,
as was also inferred for the final autocorrelation peak at 620 days in
Q0957 (S05).

We do not consider that the case for the existence of the 2 last peaks is
compelling, and we agree that the poor sampling of the light curve at the
time of the first peak adds significant uncertainty to the calculation of
inclination effects, which will be discussed later in this section.
Nevertheless, the fact that these data sets
produce a reasonable calculation of the Q2237 strong lensing time lags
(Vakulik et al, 2005) inspires confidence that important intrinsic quasar
brightening effects have been observed, and even if the quasar fully
cooperates, it will be at least 5 years before a comparable event can be
fully observed. For this reason application of the calculational procedure,
previously used for Q0957 (S05) to these data and the interpretation for
the Einstein Cross, Q2237, is fully justified.
The observed data appears to present evidence for the existence of a
complex
interior quasar structure associated with complex physical structures seen
in reverberation and having time scales of 3 years as measured by the
observer's clock. Correcting this time scale by (1+z) = 2.69 and
converting
to size scale at light propagation speed gives reverberation quasar
structure scale of approximately 1 light year or $10^{18}$cm. In the
following
paragraphs we will show that a full analysis of this data, including
allowance for inclination effects, gives a measurement of the distance
from the central object to the Elvis structure whose value is determined
to be $8.4 \times 10^{17}cm$.

We demonstrate next that if we adopt the series of brightness peaks as
continuum reverberations from the various internal quasar structures
as was done previously for Q0957 (S05 and SLR06),
we can calculate a self-consistent quasar picture that
determines the locations and sizes of the quasar structures, and also
determines the inclination of the quasar to the line of sight. To do this
we adopt the series of 4 equations from S05 for three unknown
parameters which imply an over-constrained problem. However because of the
large uncertainties in the adopted lags, we do not propose an elaborate
statistical analysis for the uncertainty of the derived parameters. The
adopted physical quasar model includes an Elvis (2000) outflow structure
which is luminous in the UV-optical continuum, and which effectively
produces a luminous ring above the accretion disc plane, and a second ring
below. The geometry of this structure is shown graphically in the two Q0957
references cited above and in Fig 1 of this paper. In the context of
this analysis the outer structure produces most of the quasar's UV-optical
luminosity. Referring back to the quasar brightness history published by the
OGLE III research group, we see that the event marking time zero, which is
probably an energy deposition in the quasar's corona- like central region
or perhaps at the inner edge of the accretion disc, produces a 30\% brightness
increase. This is followed by reverberation off the 4 surfaces of 25\%,
10\%, 10\%, and 5\%, so in a sense 80\% of the quasar's UV-optical 
luminosity is
apparently contributed by these quasar structures. These numbers for the
brightness peaks in the first and subsequent reverberations are nicely
confirmed by the Vakulik et al (2007) simulation of the observed
microlensing events seen in the OGLE III brightness curves. Because the
nearby location of the microlensing galaxy makes the microlensing sensitive
to the smaller interior structure and hardly sensitive to large outer
structure on time scales of available data, and because the amplitudes of
the cusp-shaped microlensing events must be caused by stars and can only
represent a fraction of the quasar luminosity, the Vakulik et al (2007)
results give a direct measure of the fraction of the quasar luminosity 
contributed by the central structure, which, taking the orientation angle
between the observer's line of sight and the quasar's polar axis angle into
account, has a radial size estimated from the simulation
as $1.9 \times 10^{15} cm$.

These Vakulik et al (2007) simulations show that this compact inner
structure contributes only half as much UV-optical continuum radiation, as
measured with the V filter of the OGLE-III monitoring program, as the large
outer structure. In other words, the inner structure contributes only 1/3
of the total quasar luminosity. In the preceding paragraph, we estimated
that the first impulse observed in the brightness stream was only about
$30\%$
of the total quasar brightness. The estimates are thus consistent. Notice
that the two estimates are independent, one directly from the observed
brightness record, while the other one is from a statistical microlensing
simulation.

For brevity we do not here repeat the equations solved for the quasar
structure and inclination; they are given, together with an illustration
showing their definition, in Schild (2005, Fig. 1). Solutions to this set of
equations were expressed in triads of $(r, \theta, \epsilon)$. Here r is the
quasar structure variable giving the radial distance between the compact
central object and the luminous surface of the Elvis structure. Variable
$\theta$ is the quasar orientation variable, expressed as the angular
distance
between the observer's line of sight and the quasar's polar axis. Angle
$\epsilon$ is the internal structure variable expressing the angle between
the radius to the luminous Elvis surface and the accretion disc plane.

As described in S05, two possible solutions occur because of the
unique geometry associated with the centrally illuminated Elvis structure.
In both cases, the first arriving reverberation is from the region of the
Elvis structure closest to the observer. Then case 2 is when the second
reverberation is from the part of the Elvis structure farthest from the
observer but on the same side of the accretion disc, whereas case 1 is
when
the second reverberation is from the nearest portion on the opposite side
of the accretion disc. It will always be true that case 1 is statistically
more probable than case 2, since there is just a low probability that the
quasar
pole is nearly aligned with the observer¹s line of sight to the quasar.
This is of course true only to the extent that the quasars have not been
selected
by some orientation-dependent criterion, like the beaming of the radio jet.

Our case 1 solution with $(r, \theta, \epsilon)$ of $(8.4 \times 10^{17}$ cm,
21 $\deg$, 10.7 $\deg$) has its luminous Elvis structure rings
$1.6 \times 10^{17}$ cm above the accretion disc plane. 
The case 2 solution with $(8.4 \times 10^{17}cm, 10.7 \deg, 21
\deg)$ features the Elvis structure
ring $3.0 \times 10^{17}$cm above (and below) the accretion disc plane. 
For both
case 1 and case 2, the angle between the observer and the pole of the
accretion disc plane is small, although the quasar was optically selected.

In time, better observations of the quasar brightness and color
fluctuations will allow discrimination between case 1 and case 2. In both
cases, the strongest (brightest) reverberation will always be from the
first arriving pulse, because this near-side emitting surface has no
absorption and has relativistic beaming of the observed reverberation.
The second arriving reverberation in case 2 should be approximately as
strong as the first, since it will be seen at a small angle $\theta$ giving
approximately the same relativistic beaming and no absorption. In case 1,
the second reverberation is off of the surface beamed away from the
observer, and is seen through substantial obscuration. These assertions
presume that the dark clouds of the putative dusty torus are contained just
beyond the Elvis outflow structures (see Fig. 1), 
which presumably shield them from the
hard emissions of the inner chromosphere and the inner edge of the
accretion disc. Recall that reverberation between the 2.2 {\micron} emission
and the UV/optical emission reported by Minezaki et al (2003) locate the
near-IR emission just outside the Elvis structured continuum emitting region.

From simple inspection of the brightness records for the Q2237 quasar, we
see that the second observed reverberation is less than half the amplitude
of the first. This tends to favor the case 1 geometry, which we will
therefore adopt for the remainder of this report. For case 1 geometry in
Q2237, the luminous surfaces of the Elvis structure are determined from
our solution to be $8.4 \times 10^{17}$cm from the compact central source; 
for the
purpose of the modeling calculations to be discussed later in the paper it
is important to know how far this is in the accretion disc plane.
Multiplying by cos 10.7 $\deg$, we find that the luminous surfaces lie
$8.2 \times 10^{17}$cm from the center and $1.6 \times 10^{17}$cm above 
and below the plane.
The opening angle (the angle between the observer's line of sight and the
jet outflow, with the central compact source at the vertex) is 79.3
degrees.

Finally we note that for the solution of the structure equations proposed
above, we can predict the arrival time of the final brightness peak. In
either case 1 or case 2 orientation, the final peak should arrive 1332
days (observer's clock) following the initial outburst.

\section{Image Brightness Enhancement Corrections to the Observed 
Fluxes From Q2237 and Q0957}

Because the multiple images associated with gravitational lenses are usually
separated by only a few arc seconds in the plane of the sky, historically
observations of them at some wavelengths require correction for the number
of images if the intrinsic quasar luminosity is to be inferred from a
brightness measurement of the combined image. Furthermore, the lensed
images
are magnified by the lensing, so determination of the intrinsic quasar
fluxes requires a correction for the magnification. Accurate knowledge of
the fluxes is important to our analytical model of the quasar structure as
summarized in Tables 1 and 2. Then as discussed in section 4 below, if the
distantly observed MECO period of rotation can be estimated from
observations of the outer light cylinder radius of the MECO, the 
analytical model then allows us to
estimate the MECO mass, accretion rate, and magnetic field strength (with
rotation rate estimated from the light cylinder radius)

\subsection{Image Brightness Corrections to the Observed Flux of Q2237}

For Q2237 in the radio quiet high soft state, four images are seen and so
the published total radio flux estimate for the entire system must be
divided by four. While the magnification factor is still uncertain, the
most recent models by Schmidt and Wambsganss (2004) suggest a magnification
factor of sixteen. Early radio observations of the Einstein Cross concluded
that the quasar was radio quiet, but a more heroic 11 hour integration with
the VLA in A configuration produced a 3.6 cm total flux measurement of 593
+/- 88 micro-Janskys (Falco et al 1998). With correction of factor 4 for the
image multiplicity and 16 for strong lensing magnification, the inferred
quasar 3.6 cm radio flux is 9.3 micro-Janskys, which puts it in the class
of radio quiet quasars.

The X-ray luminosity of Q2237 has been measured in two Chandra observations
over the 0.4 - 8.0 Kev band reported by Dai et al (2004). Their uncorrected
total flux measurement of $9.2 \times 10^{45}$ergs/sec must be corrected 
by 4 for
the image multiplicity and by factor 16 for strong lensing; we adopt a
corrected quasar X-ray luminosity of $1.4 \times 10^{44}$erg/sec. An
important discovery of the Chandra observation of Q2237 was the observation
of a red shifted broad $FeK_{\alpha}$ line at 5.7 KeV. 
The estimated FWHM is 0.87
KeV and equivalent width is 1200 eV. Thus for a rest wavelength of 8.7 KeV
the red shift is 0.52 (velocity $1.6 \times 10^{10}$cm/sec) and 
the velocity width
is $0.45 \times 10^{10}$cm/sec (45,000 km/sec). Because the line 
is prominent in
image A and unseen or weak in images B, C, and D, it must be microlensed, as
discussed by Popovic et al(2006), who conclude that the region emitting the
emission line must be smaller than the continuum emitting region,
consistent with the suggestion that the line emission comes from the thin
inner edge of the accretion disc. Popovic et al also noticed that because
of the short duration and strength of the microlensing event, micro lenses 
with mass on the order of a planetary mass are required if the central
compact quasar mass is on the order of $3 \times 10^{9}M_\odot$, 
as estimated here.

\subsection{Image Brightness Corrections to the Observed Flux of Q0957}

For Q0957 in a radio loud low hard state, two images are seen and the
total flux estimate for the entire system must be divided by two. While
the magnification factor is still uncertain the most recent models by
Schild \& Vakulik (2003), which include the microlensing magnification,
suggest an overall brightness amplification factor of three.

Radio observations of the Q0957 have shown that the quasar is radio loud
and
has a jet whose radio brightness is dominated by the core of the jet. Using
a correction of factor 2 for the image multiplicity and a factor 3 for
strong lensing magnification. The inferred quasar  6 cm radio flux  has
been
determined to be 9.3 milli-Janskys (Lehar et al 1991, Schild (2005) and
Schild, Leiter, and Robertson (2006)) giving Q0957 a radio luminosity of
$3.7 \times 10^{42}$erg/sec.

The X-ray luminosity of Q0957 has been measured in Chandra observations
over
the 0.4 - 8.0 Kev band reported by Chartas et al (2001). Their
uncorrected total flux measurement of $1.5 \times 10^{46}$ergs/sec must be
corrected
by 2 for the image multiplicity and by factor 3 for strong lensing; we
adopt
a corrected quasar luminosity of $2.5 \times 10^{45}$erg/sec. We also 
note that an
important conclusion of the Chandra observation of Q0957 was the apparent 
lack of any red shifted broad Fe-Kalpha line at 5.7 KeV.

\section{Analysis of the Internal Structure of Q2237 and Q0957 in Different
Spectral States}

In the following sections we will compare the observed inner structure for
Q2237 (the Einstein Cross quasar in the radio quiet high soft spectral
state) to the inner structure of Q0957 (the Twin quasar in the radio loud
low hard spectral state) which was previously established in the 2006
paper
by Schild, Leiter, and Robertson. In the process we will find that the
explanation of the physical difference between the two quasars Q0957 and
Q2237 is related to the more general question as to why some quasars are
observed to be radio loud and have a hard X-ray spectrum while others are
observed to be radio quiet and have a soft X-ray spectrum. In this section
we will compare and analyze the data for these two quasars, as shown in
Tables 1 and 2 below, in terms of a new theoretical quasar model which is
easily
able to explain the physical nature of these quasars in their different
spectral states as well as providing new understanding that can bring
unification to a wide body of quasar data.

\subsection{Analysis of the Internal Structure of Quasar Q0957 in the Radio
Loud Low Hard State (RL-LHS)}

We begin by recalling that in SLR06, we examined the empirical data for the
lensed and micro-lensed Q0957+561A,B quasar obtained from 20 years of
brightness monitoring at visible wavelengths (near-ultraviolet emission at
the quasar). In that report we examined several conclusions inferred
previously from analysis of the auto-correlation and micro-lensing
properties of the monitoring data, and collected these results in a
consistent presentation and used them to confront physical quasar models
and their simulations. On the basis of the analysis in SLR06 the inner
structure
observed for the quasar Q0957 in the (RL-LHS), which we referred to as the
Schild-Vakulik Structure, was made up of four component parts which had
the following dimensional properties (see figure 1):

\begin{itemize}
\item 1. Elliptical Elvis Structure:  $Re = 2 \times 10^{17}$cm, He = $5
\times 10^{16}$cm 
   
\item 2. Inner Radius of Accretion Disk: $R_{disk} = 4 \times 10^{16}$cm
   
\item 3. Hot Inner Accretion Disk Annulus:  delta(R) = $5.4 \times 10^{14}$cm
   
\item 4. Base of Radio Jet: $R_{rad} = 2 \times 10^{16}$cm , $H_{rad} = 9
\times 10^{16}$cm 

\end{itemize}
The inner components of this quasar structure were found to be an accretion
disc truncated at a large radius whose hot inner edge contained a bright 
thin annulus. This inner structure was found to be surrounded by an
elliptical coronal outflow (Elvis Structure) long known to explain the
complex spectroscopic behavior observed in quasars. However it was observed
that the opening angle of the coronal Elvis structure with respect to the
z-axis of rotation appeared to have a very large value of 76 degrees. In
addition a radio emission region was found to be located directly above the
compact source. The size and location of this radio emitting region was
found to be where the reconnection of magnetic field lines at relativistic
Alfven speeds, like that generated by a rotating central object containing
an intrinsic magnetic field, should occur. Hence the structure observed in
the quasar Q0957+561 appeared to closely resemble the complex
inflow-outflow
patterns seen in intrinsic magnetic propeller models for young stellar
objects (Romanova et al, 2002, 2003a, 2003b, 2004).

In SLR06 it was shown how standard black hole models where unable to account
for the internal structure observed in the quasar Q0957.  In particular we
discussed how attempts to model the observed internal structure seen in the
quasar Q0957+561 in terms of the intrinsic magnetic moment generated by a
central spinning charged black hole failed because the necessary charge on
the spinning black hole required to make it work would not be stable
enough
to account for the long lifetime of the observed Schild-Vakulik structure. 
Similarly it was
shown how attempts to model the internal Q0957 structure in terms of Kerr
Black Hole-ADAF-Accretion Disc Corona-Jet Models, in which the magnetic
field is tied to the accretion disk and not the central rotating compact
object, failed in that this model was unable to account for the very large
opening angles observed for the coronal Elvis outflows. Finally it was
shown
that Magnetically Arrested Disc (MAD) black hole models failed in that
they
predicted the existence of orbiting infalling hot blobs of plasma inside
of
the inner region of the accretion disk that were not observed.

On the other hand it was shown in SLR06 that the four components of the
intrinsically magnetic structure observed in Q0957 could be successfully
explained by Magnetospheric Eternally Collapsing Object (MECO) models
(Robertson and Leiter 2003, 2004, 2005), which feature highly redshifted,
Eddington limited, collapsing central compact objects containing a strong
intrinsic magnetic moment aligned with the MECO axis of rotation (see
Appendix 1 for more details). This is indicated in Table 1 and Table 2
where, taking into account the amplification factor of three discussed in
section 3.b it is shown that the MECO model for the quasar Q0957+561 can
be represented by a $4.2 \times 10^{9}M_\odot$ mass scaled up version of 
the class of MECO
models that have been used to explain "spectral state switching" phenomena
associated with Galactic Black Hole Candidates. Such a massive MECO object
does not collapse beyond its event horizon. Instead, in-falling matter
heats up and because if its intense radiation collects at a highly redshifted
physical surface which is balanced in a state of local Eddington limited
secular equilibrium at a radial distance which is just outside of its event
horizon. This allows the highly redshifted MECO surface 
to slowly collapse, while remaining in causal connection
outside of its event horizon, 
over time intervals greater than a Hubble time as
seen by a distant observer.
Because of the small light cone opening angle for radiation escaping
from this highly red shifted MECO surface to a distant observer, the resultant
low luminosity in the far-infrared wavelengths make this MECO surface 
difficult to detect for the case of Q0957+561.

In SLR06 it was demonstrated that Q0957 could be successfully modeled only
as a MECO driven quasar in the Low Hard State in which the MECO intrinsic
magnetic propeller interacted strongly with the inner regions of the
quasar's accretion disc. For Q0957 in the Low Hard State, the dynamo
action of the intrinsic rotating MECO magnetic dipole field of its central
supermassive MECO sweeps clean the central region of the quasar out to the
magnetospheric radius where the magnetic propeller acting on the inner
edge of the accretion disc creates a very thin annular band, and a radio
emitting
region above the disc where the magnetic field lines twist and bunch up
until they eventually break and reconnect at relativistic speeds. The
resultant size and location of the radio emitting region associated with
this inner magnetic structure of the quasar was found to be in the
region above the central supermassive MECO object where the reconnection
of magnetic field lines at relativistic Alfven speeds, like that generated
by a rotating central MECO containing an intrinsic magnetic field, should
occur.

For the quasar Q0957 the outer bi-conic Elliptical Elvis Coronal Outflow
Structure, that was observed to occur with a very wide opening angle above
and below the accretion disk at a distance 
$R_{Elvis} \sim 2 \times 10^{17}$cm, was found
to be located just inside of the outer light cylinder of its central MECO
which occurred at a radial distance of $R_{lc} \sim 2.4 \times 10^{17}$cm. 
See Table 1-2
where this observation, and the formula for the MECO outer light cylinder
given by Rlc $\sim (1.5 \times 10^{17}cm) \times T_{year}$, was 
used to estimate the distantly
observed rotation period of the central MECO in Q0957 as 
$T_{Year} \sim 1.6$ years.

The existence and the radial location of bi-conic Elliptical Elvis Coronal
Outflow Structures, occurring just inside of the outer 
light cylinders of
central compact MECO magnetic objects surrounded by hot accretion disks, is
a predicted observational property of the MECO-quasar accretion disk model
that uniquely distinguishes it from a black hole-quasar accretion disk
model.

The MECO-quasar model satisfies the physical conditions required to
generate
bi-conic shaped Elvis Outflow Structures just inside of their outer light
cylinders by virtue of the combined action of two well-known physical
processes:
    a) Magnetic reconnection effects which can generate upward flows just
inside of the outer light cylinder of central compact magnetic objects
surrounded
by accretion disks (Uzdensky, 2003, 2006; Uzdensky; Spitovsky, 2006),
and
    b) Optical line driving forces, associated with the presence of strong
ultraviolet luminosity emitted from the inner magnetospheric radius of a
hot accretion disk surrounding the central compact magnetic object, which
can act on the magnetic reconnection generated upward flows just inside of
the outer light cylinder and there create the bi-conic Elvis Outflow 
Structures (Elvis 2000, 2003, 2006).

In this manner the intrinsic magnetic field of the MECO generates a vertical
up-flow of plasma above the accretion disk by magnetic reconnection effects
which occur just inside of the outer light cylinder of the MECO. The
vertical up-flow of plasma occurs where the surrounding dusty torus ends
and the bi-conic Elvis Outflow Structure begins (see Fig. 1). If the inner
accretion disk is hot enough to make UV radiation then the effect of
photo-ionization line-driving bends the upward flowing plasma outward over
the accretion disk and generates the bi-conic Elvis Outflow Structure of
positive ions associated with the Broad Emission Line Region (BELR). Charge
separation between the electrons and outflowing ions is generated by the
photo-ionization process near the MECO light cylinder. The return current
of ions most likely flows back to the central region along the separatrix
associated with field-line opening due to the differential rotation which
occurs in the magnetically-linked MECO ­ disk system.

\subsection{Analysis of the Internal Structure of Quasar Q2237 in the 
Radio Quiet High Soft State}

As discussed in sections 3.a the detailed internal structure of quasar 
Q2237 in a radio quiet high-soft state has been observed to be as follows:
\begin{itemize}
\item 1. Elliptical Elvis Structure: $Re = 8.2 \times 10^{17}$cm , 
$He = 5 \times 10^{16}$cm 

\item 2. Inner Radius of Accretion Disk:  $Rdisk = 1.9 \times 10^{15}$cm

\item 3. Broad Iron Line Emission seen:  Yes

\item 4. Base of Radio Jet: None since there is no jet in Radio Quiet State

\end{itemize}

In the context of the MECO model the internal structure of the quasar Q2237
can be easily understood in terms of different spectral states (i.e. Radio
Loud Low Hard States versus Radio Quiet High Soft States) which are
distinguished by the nature of the interaction between the accretion disks
in these quasars and the intrinsic rotating magnetic moments of their
central supermassive MECO. We begin by recalling that in section 4a we
found that for Radio Loud Low Hard States like that seen in Q0957, the
magnetospheric radius Rm is greater than its MECO co-rotation radius Rcr and
the resultant magnetic propeller effect on the accretion disk generates an
empty inner region inside of the accretion disk, which is surrounded by a
thin hot ring, and a jet outflow leading to strong radio emission. However
for Radio Quiet High Soft States (RQ-HSS) like that observed in Q2237, the
accretion rate is high enough to push the magnetospheric radius Rm inside
of its MECO co-rotation radius $R_{cr}$. This shuts down the 
magnetic propeller, and
the jet outflow which creates a spectral change from a RL-LHS to a RQ-HSS
state. In this state the intrinsic magnetic propeller mechanism of the
MECO-AGN becomes incapable of ejecting the flow in the form of a jet, which
causes an increasingly dense boundary layer of plasma to form at the inner
disk radius which eventually pushes it in all the way to the marginally
stable orbit. When this occurs the MECO-AGN changes its spectral state from
a Radio Loud Low Hard State (RL-LHS) to a Radio Quiet High Soft State
(RQ-HSS).

To get a better physical insight into the nature of this process let us
discuss the dynamics of the MECO-quasar spectral state change from RL-LHS
state to the RQ-HSS state in a little more detail. For the higher accretion
rates associated with the RQ-HSS the intrinsically rotating MECO magnetic
field is dynamically unable to eject the disk material from the interior
region inside of the co-rotation radius $R_{cr}$, 
and the accreted matter piles up
against the magnetopause and pushes it inward. The fact that a boundary layer
at the inner accretion disk radius must exist around a MECO in the (RQ-HSS)
was demonstrated (Robertson and Leiter 2005 section 11, pp 24-25) by
comparing the magnetic pressure at the magnetosphere with the impact
pressure of a trailing, subsonic disk where it was shown that radial inflow
disk velocities in excess of the speed of light would be required to allow
the impact pressure to match the magnetic pressure. While the gas pressure
in the disk at the inner radius of the transition zone is normally
required
to match the magnetic pressure in the disk, it turns out that for the
(RQ-HSS) state the radiation pressure in the disk is less than the gas 
pressure. Hence for a MECO-AGN in the RQ-HSS the boundary layer, located
at the inner disk radius of its gas pressure dominated, thin, subsonic
Keplerian disk, will eventually be pushed down to the marginally stable
orbit of its accretion disk so that $R_{disk} \sim R_{ms}$. 
The value of $R_{ms}$ will
depend on the a/M ratio of the rotating Vaidya metric which describes the
MECO (which, because of the very highly redshift of the MECO, will to a
good
approximation be similar to the value of $R_{ms}$ for a Kerr metric). In
order to apply the above analysis to the case of Q2237 in the RQ-HSS we
recall from section 3.a that Vakulik et al (2007) found that the 
fraction of the
quasar luminosity contributed by the central structure was associated with a
radial size of $1.9 \times 10^{15}$cm. 
Assuming that this is the inner radius of the luminous accretion disc edge
for the case of Q2237 in the RQ-HSS, we find that $R_{disc}$ $\sim$ $R_m$ 
$\sim$ $1.9 \times 10^{15}cm = 4.2 R_{grav}$.
This is consistent with the central MECO in Q2237 having a mass of $3
\times 10^9 M_{odot}$ and a specific angular momentum ratio of a/M = 0.5.
In this context we find that
the observed core X-ray luminosity of $Lx \sim 1.4 \times 10^{44}$erg/sec, 
which in the
RQ-HSS is generated in the corona above the accretion disk, will interact
with the inner region of the accretion disk and generate broad Iron
emission lines consistent with Q2237 observation.

In addition we find that the properties of MECO-quasar model for Q2237
satisfy the physical conditions required to generate bi-conic shaped Elvis
Outflow Structures just inside of their outer light cylinders by virtue of the
combined action of magnetic reconnection effects (which generate upward
flows just inside of the light cylinder of central compact magnetic
objects
surrounded by accretion disks) and optical line driving forces (associated
with the presence of strong ultraviolet luminosity emitted from the inner
magnetospheric radius of a hot accretion disk surrounding the central
compact magnetic object) which act on the magnetic reconnection generated
upward flows that occur just inside of the outer light cylinder and 
there create the bi-conic
Elvis Outflow Structure observed. 
For the case of the quasar Q2237 the outer bi-conic
Elvis Outflow Structure, with a very wide opening angle,
which was observed above and below the accretion disk at a distance
$R_{Elvis} \sim
8.2 \times 10^{17}$cm, is found to be located just inside the outer light
cylinder
of its central MECO at $R_{lc} \sim 8.4 \times 10^{17}$cm (note that 
this result is
obtained from $R_{lc} \sim (1.5\times 10^{17}cm) \times T_{year}$, where $T_{year}$ is the distantly
observed rotation period of the central MECO, which for the case of Q2237
was found to be $T_{year} \sim 5.6$ years). It is important to note that the
existence of bi-conic Elliptical Elvis Coronal Outflow Structures, predicted
to occur just inside of the outer light cylinders of central compact MECO
magnetic objects surrounded by hot accretion disks, is a key observational
prediction of the MECO-quasar accretion disk models which distinguishes it
from black hole-quasar accretion disk models.

\section{CONCLUSIONS}

We have shown how direct microlensing-reverberation analysis performed on
two well-known Quasars (Q2237 - The Einstein Cross and Q0957 - The Twin)
can be used to determine the inner structure of two quasars which are in
significantly different spectral states. These observations allowed us to
measure the detailed internal structure of quasar Q2237 in a radio quiet
high-soft state, and compare it to quasar Q0957 in a radio loud low-hard
state. We found that a unified explanation of the dynamics of the inner and
outer components of the observed internal structure of these two quasars
could be dynamically explained as being due to different spectral states
(i.e. Q0957 in a Radio Loud Low Hard State (RL-LHS), and Q2237 in a Radio
Quiet High Soft State (RQ-HSS)). We then argued that when taken together,
the most plausible physical explanation for the observed differences in
the
spectral states of these two quasars was due to the location of the inner
radii of their accretion disks (see Fig. 2) relative to the 
co-rotation radii associated
with supermassive, rotating "Magnetospheric Eternally Collapsing Objects"
(MECO) in the centers of these quasars (ref Appendix I).

In support of the above argument we first described the unique
gravitational
microlensing observations of the quasar Q0957 in the (RL-LHS) which
permitted a reconstruction of its internal radiation emitting structures
(SLR06). In the context of these observations the inner accretion disk
was found to be truncated at a very large radius and surrounded by a very
thin hot inner ring of plasma with a large empty inner region inside of it.
In addition large bi-conic Elvis outflow structures were observed about ten
times further out from the inner disk radius. The overall morphology of
these dynamic structures were found to be similar to features revealed in
simulated accretion flows into compact Young Stellar Objects which
contain a
central rotating magnetic dipole objects inside of their accretion disks
(Romanova 2002, 2003a). Since black holes do not contain observable intrinsic
dipole fields, such dynamic structures cannot be generated by black
hole-accretion disk models. Hence our conclusion was that we were
observing
the physical effects of a central supermassive compact intrinsically
magnetic object in Q0957, and that instead of a black hole the
super-massive
compact object in this quasar was a MECO

Then applying the MECO model to the case of Q2237 in the (RQ-HSS), we argued
that the higher accretion rates involved pushed the inner magnetospheric
radius of its accretion disk through the co-rotation radius of its central
MECO and shut down the MECO magnetic propeller effect on the accretion disk.
Since the lack of magnetic propeller action on the accretion disk turned off
the radio-jet emission leading to the creation of a radio quiet state, and
caused the inner radius of the accretion disk to be pushed down to the
marginally stable orbit, then X-rays generated in the corona above the
accretion disk were able to act on the inner edge of the accretion disk and
create the broad iron line 9.7 Kev signature observed in Q2237. In support
of this idea we note that models of the inner region surrounding the inner
accretion disc edge, for young stellar objects by Romanova (2002, 2003a,b;
2004) and
others, feature an inner corona of ionized matter above the accretion
disk. 

Detailed calculations from analytical modeling shown in Table 1-2, showed
how the MECO model can quantitatively explain the inner dynamic structures
of both the quasar Q0957 in a RL-LHS state as well as that of the quasar
Q2237 in a RQ-HSS state. In the MECO model the different spectral states
of these two quasars were determined by the magnitude of their respective
accretion rates and the interaction between their accretion disks and the
intrinsic rotating magnetic moments of their central supermassive MECO. In
addition the MECO model for Q0957 and Q2237 connected the location of their
Elvis outflow structures to the location of the outer light cylinders of 
their central rotating supermassive MECO.

The MECO model of internal quasar structure presented here, for Q0957 in
the radio loud low hard state and Q2237 in the radio quiet high soft
state,
can be readily applied to a wide body of data on other quasars observed in
one or the other of these two spectral states and for this reason has the
potential to be able to unify a vast amount of quasar spectroscopic 
and photometric data.

In addition the MECO quasar model has the unique property of dynamically
connecting the location of the inner part of the quasar Elvis outflow
structure with the location of the outer light cylinders of the central
MECO
in these quasars. Because of the special geometric relationship that
elliptical Elvis outflow structures have to the modified dusty torus
components of their Elvis structures as shown in figure 1, the MECO quasar
model predicts that for any orientation of the quasar to the line of
sight,
some component of emission should be evident for any optical depth of
dusty
torus absorption. This result is based on the assumption that the
continuum
emission originates principally in a corona near the inner edge of the
accretion disc, and at the outflow structures illuminated by the inner
corona and accretion disc. This is because in MECO quasar model, the Elvis
structures form the inner edge of the dusty torus.  Then for all viewing
angles the UV-optical power-law continuum should be visible, except
possibly
for extreme equator-on geometry (with the pole of the accretion disc
perpendicular to the line of sight.)

In summary we have demonstrated how direct microlensing-reverberation
analysis of two Quasars (Q2237 - The Einstein Cross and Q0957 - The Twin)
have allowed us to observe inner structure of two quasars which are in
significantly different spectral states. Standard black hole accretion
disk
models were found to be unable to offer a unified dynamical explanation of
the observed internal structure of both of these two quasars in different
spectral states.  Instead we found that a unified explanation could be
obtained within the framework of a MECO model. This was done by describing 
the spectral states of the
two quasars in terms of the location of the magnetospheric radii of their
accretion disks with respect to the co-rotation radii, inside of the light
cylinder surrounding highly red shifted, rotating $M \sim 10^{9}M_{\odot}$ 
MECO in the centers of these quasars.

Because of the apparent ease with which the size of the Elvis structure and
the quasar orientation with respect to the plane of the sky can be
inferred, following the method of S05, future photometric
observations with the planned Large Scale Synoptic Telescope (LSST) will
allow an important MECO parameter to be easily obtained for any of the
thousands of quasars to be monitored. Since the radius of the MECO outer
light cylinder depends only on the distantly observed MECO period of
rotation, the connection between the MECO outer light cylinder and the
location of the Elvis outflow structures observationally determines the
distantly observed MECO spin parameter. Hence, in addition to the MECO mass
and intrinsic magnetic moment, simple recognition of patterns of
reverberations and their timing allows distantly observed MECO spin
parameters to be obtained.

\section{APPENDIX I. THE STRONG PRINCIPLE OF EQUIVALENCE AND MAGNETOSPHERIC
ETERNALLY COLLAPSING OBJECTS (MECO)}

In General Relativity, preservation of the Strong Principle of Equivalence
(SPOE) requires that Special Relativity must hold locally for all
time-like
observers in all of space-time. The existence of MECO is implied by the idea
that Nature requires that the Strong Principle of Equivalence (SPOE) be
dynamically preserved everywhere in space-time for the timelike world lines
of massive particles or fluids under the influence of both gravitational and
non-gravitational forces. Preservation of the SPOE requires that the frame
of reference of the co-moving observer in the massive collapsing fluid must
always be connected to the frame of reference of a stationary
observer by special relativistic transformations with a physical 3-speed
that is less than the speed of light (SLR06 and Landau \&
Lifshitz, 1975). Since the left hand side of the Einstein equation cannot by
itself enforce the dynamic preservation of the SPOE then SPOE-preserving
non-gravitational processes must exist in Nature and must always be included
in the energy-momentum tensor on the right hand side of the Einstein
equation.

It was in this context that the general relativistic MECO solutions to the
Einstein-Maxwell equations were discovered, as was shown in the three
previously published papers of Robertson and Leiter and developed in more
detail in Appendix 1-10 of Astro-ph/0505518. There it was shown that for a
collapsing body, the structure and radiation transfer properties of the
energy-momentum tensor on the right hand side of the Einstein field
equations, could describe a collapsing radiating object which contained
equipartition magnetic fields that generated a highly redshifted Eddington
limited secular collapse process. This collapse process was shown to
preserve the SPOE by dynamically preventing trapped surfaces, that lead to
event horizons, from forming.

In Appendix 1-10 of Astro-ph/0505518 it was shown that, by using the
Einstein-Maxwell Equations and Quantum Electrodynamics in the context of
General Relativistic plasma astrophysics, it was possible to virtually
stop
and maintain a slow, (many Hubble times!) steady collapse of a compact
physical plasma object outside of its Schwarzschild radius.

The non-gravitational force was Compton photon pressure generated by
synchrotron radiation from an intrinsic equipartition magnetic dipole
field
contained within the compact object. The rate of collapse is controlled by
radiation at the local Eddington limit, but from a highly redshifted
surface.

In Appendix 9 and 10 of Astro-ph/0505518, it was shown that general
relativistic surface drift currents within a pair plasma at the MECO
surface
can generate the required magnetic fields. In particular it was shown in
Appendix 9 that the equatorial poloidal magnetic field associated with 
a locally Eddington limited secular
rate of collapse of the exterior surface was shown to be strong enough to
spontaneously create bound electron-positron pairs in the surface plasma of
the MECO. In the context of the MECO highly redshifted Eddington limited
balance, the action of this QED process was shown to be sufficient to
stabilize the collapse rate of the MECO surface.

For the case of hot collapsing radiating matter associated with the MECO,
the corresponding exterior solution to the Einstein equation has been
shown
to be described by the time dependent Vaidya metric where no coordinate
transformation between MECO Vaidya metric and the Black Hole Kerr-Schild
metric exists. Since the highly redshifted MECO Vaidya metric solutions
preserve the SPOE they do not have event horizons, and in addition they
exhibit a distantly observed slowly rotating intrinsic magnetic dipole
moment which interacts with a surrounding accretion disk. Hence the MECO
magnetic moments will have observable effects which distinguish them from
Black Holes if such MECO exist at the centers of galactic black hole
candidates and AGN.

ACKNOWLEDGEMENTS

The authors wish to thank Alan Bridle, of the National Radio Astronomy
Observatory (NRAO) in Charlottesville Virginia, for many important
conversations and analytical discussions which played a key role in
clarifying the ideas presented in this paper. In particular one of us (DL)
thanks him for graciously allowing access to the research facilities of
NRAO during the important developmental phase of this work.

\begin{table*}
\begin{center}
\caption{MECO-QUASAR MAGNETOSPHERIC EQUATIONS - (RL-LHS) AND (RQ-HSS)
- [ref. Astr J. v132,420,(2006) and Astro-ph/0505518]}
\end{center}
\begin{tabular}{lll} \hline
MECO Physical Quantity & Equation & Scale\\
\hline
Surface Redshift - & $1+z_s = 1.5\times 10^8 ({M/7})^{1/2}$ & $M^{1/2}~$\\
Surface Luminosity - & $L_s=L_{Edd}/(1+z_s)$ erg/s & $M^{1/2}~$ \\
Surface Temp. - & $T_s=2.3\times 10^7/[M(1+z_s)]^{1/4}$ K & $M^{-3/8}$\\
Rotation Rate (1/Tyear) - & $\nu=2.8[L_{q,32}/M]^{0.763}/L_{c,44}$cyl/year
& $M^{-1}~$\\
Quiescent Lum. - & $L_{q,32}=0.65 M [L_{c,36}/M]^{1.75}$ erg/s & $M$~~\\
Co-rotation Radius - & $R_c=46.7 R_g/[M \nu_2]^{2/3}$ cm & $M$~~\\
Light Cylinder Radius - & $R_{lc}=[1.52(Tyear)\times 10^{17}]$ cm &$M$~~\\
Magnetic Moment - & $\mu_{27} = 8.16[L_{c,36} M/\nu_2^3]^{1/2}$
$$gauss-cm$^3$ & $M^{5/2}~$\\
Magnetic Field - & $B_\theta = 1.12
(R_g/r)^3[L_{c,36}/(M^5\nu_2^3)]^{1/2}$ gauss
& $M^{-1/2}$\\
Magnetosphere Radius (RL-LHS) - & $R_m= 17R_g[M
\nu_{2}]^{4/21}/(L_{disk}/L_{Edd})^{2/7}$ cm & $M$~~\\
Magnetosphere Radius (RQ-HS) - & $R_m = R_{ms} $ &
$M$~~\\
Core X-ray Luminosity - & $L_{x}= (\eta_{x-corona-disk})L_{disk}$ erg/s
&$$~
\\
Core Radio Luminosity - & $L_{rad,36}=C_{rad-x} M^{0.75}
L_{x,36}^{2/3} {{|1-{(L_{x,36}/L_{c,36})}^{1/3}|}}$ erg/s &
$M^{3/2}$~ \\
\hline
\end{tabular}\\

In Table 1:

      a)  the left hand column summarizes the MECO physical quantities
relevant to Q0957 and Q2237.

      b)  the middle column gives the specific form of the equations to be
used in the calculation of these specific physical quantities and;

      c)  the right hand column gives the functional form of the relevant
mass scaling of these specific physical quantities in order to demonstrate
clearly how the equations for these physical quantities can be applied to
the case of both GBHC, AGN and QSO.

Then using the observed values of the X-ray luminosity, radio luminosity,
red shift and the physical dimensions of the internal structure seen in
Q0957 and Q2237 as input in the MECO equations in Table 1, the results
shown in Table 2 are obtained.
\end{table*}
\newpage

\begin{table*}
\begin{center}
\caption {QUASAR Q2237 (RQ-HSS) VERSUS QUASAR Q0957 (RL-LHS)}
\end{center}
\begin{tabular}{lcc} \hline
MECO Parameters & QUASAR 2237 (RQ-HSS) & QUASAR Q0957 (RL-LHS) \\
\hline
MECO Mass & $3.0\times 10^9 M_\odot$ & $4.2\times 10^9 M_\odot$\\
MECO Magnetic Moment & $7.8\times 10^{51}$ gauss-cm$^3$ & $1.8\times
10^{52}$ gauss-cm$^3$\\
MECO Magnetic Field & $(4.0\times 10^5$ gauss)($3Rg / r)^3$ & $(3.4\times
10^5$ gauss) $(3Rg / r)^3$\\
MECO a/M Ratio & $\sim 0.5$ & $>0.7$\\
MECO Surface Redshift & $3.1\times 10^{12}$ & $3.7\times 10^{12}$ \\
MECO Rotation Period - Tyears& $5.6$ Years & $1.6$ Years\\
MECO Surface Luminosity & $1.3\times 10^{35}$ erg/sec & $1.5\times10^{35}$
erg/sec\\
MECO Surface Temp & $74$ K & 66 K\\
MECO Co-rotation Radius & $151 R_g$ & $53 R_g$\\
MECO Light Cylinder-cm &$R=1867 R_g = 8.4\times 10^{17}$ & $R=383 R_g =
2.4\times 10^{17}$ \\
Accretion Disk Luminosity & $6.2 \times 10^{45}$ erg/sec & $4.6\times
10^{45}$ erg/sec\\
Eddington Ratio & $0.02$ & $9.0\times 10^{-3}$ \\
Core X-ray Luminosity (OBS) & $1.4 \times 10^{44}$ erg/sec & $2.5\times
10^{45}$ erg/sec\\
Core Radio Luminosity (OBS) & $1.0\times 10^{40}$ erg/sec & $3.7\times
10^{42}$ erg/sec\\
Core X-ray Efficiency & $\eta_{x-corona-disk}=0.03$ &
$\eta_{x-corona-disk}=0.54$ \\
Core Radio X-ray Luminosity Coeff. & $C_{rad-x}=2.8\times 10^{-9}$ &
$C_{rad-x}=3.5\times 10^{-7}$ \\
Rm Magnetosphere-cm (OBS) &$R_{m} = 4.2R_g = 1.9\times 10^{15}$ & $R_m = 64
R_g = 4\times 10^{16}$ \\
Rm Emission Wavelength (OBS) &$2000 A_o$ & $2697 A_o$ \\
Hot Annular Ring at Rm-cm (OBS) &$\delta(R_m)=0.025 R_g=1\times10^{13}cm$& 
$\delta(R_m) = 1.0 R_g
=5.4\times10^{14}$\\
Base of Radio Jet-cm (OBS) &$$ NONE-COMPACT SOURCE & $R=2\times 10^{16}$
$H=9\times 10^{16}$ \\
Elvis Outflow Structure-cm (OBS) &$R=8.2\times 10^{17}$ $H=1.6\times
10^{17}$ & $R=2\times 10^{17}$ $H=5\times 10^{16}$\\
\hline
\end{tabular}\\
\bigskip

In Table 2 the data presented is intended to be read as follows:

The scaling relations from Table 1 and other
input data are used to determine the properties of Q2237 in the Radio
Quiet High Soft State and Q0957 in the Radio Loud Low Hard State.
We emphasize in the lower rows the model derived predicted values of
observable parameters, principally the observable size scales. However many
of these size scales have already entered the MECO model calculations, but
are repeated here for a complete listing of observable and physical
properties. 
\end{table*}

\newpage
\begin{figure}
\begin{center}
\plotone{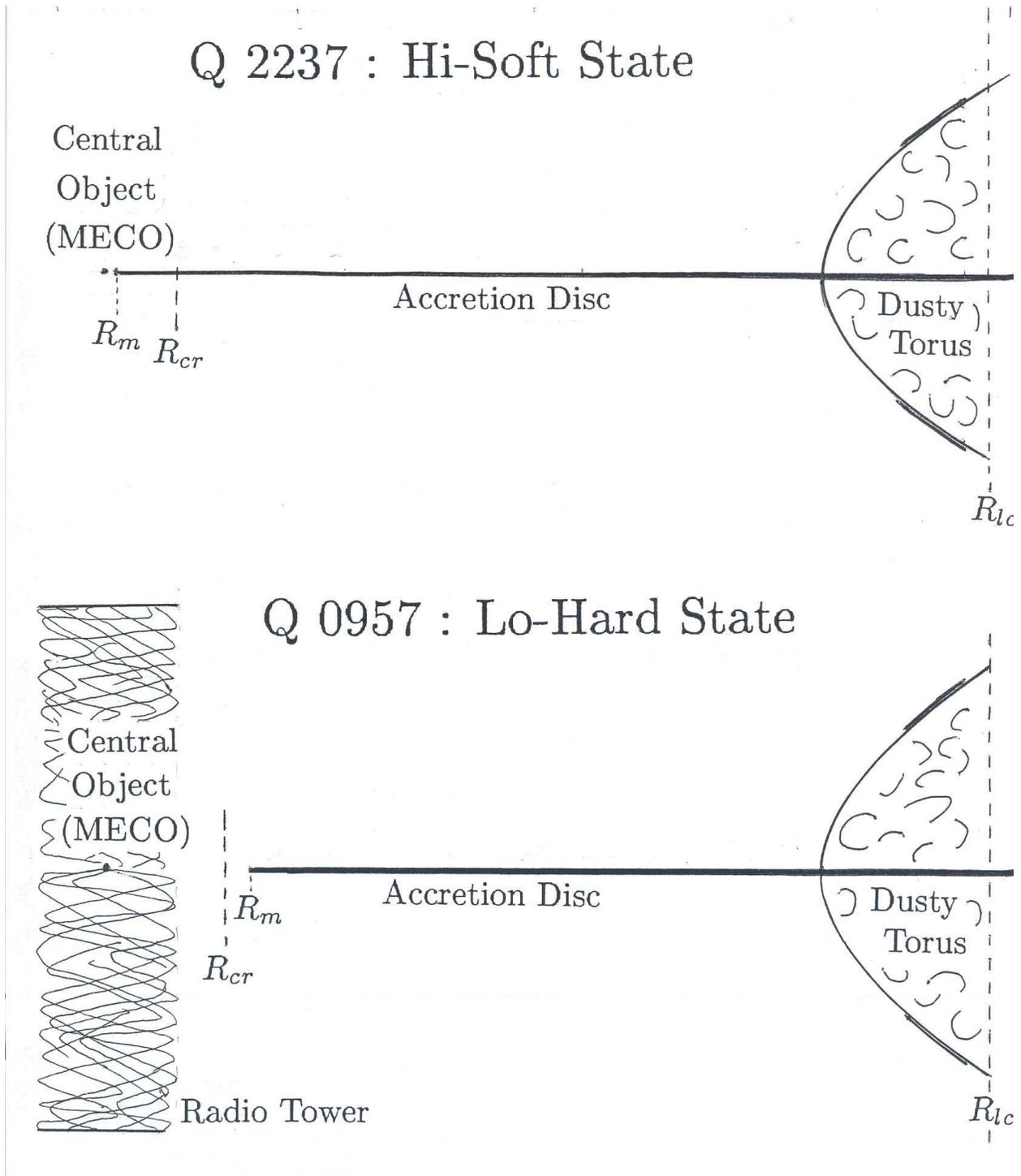}
\caption{A cross-section view comparing the structures of two quasars with
structure determined by microlensing-reverberation analysis. For both
quasars, the compact central object (MECO) is a small dot on the left, and
to its right is the accretion disc plane for the right-hand side of the
quasar. To aid comparison of the visible structures, both have been scaled
to make the distance from the central object to the light cylinder and
Elvis structure the same, although in reality the Q2237 quasar is $\times
3.5$ larger. Labels show the locations of the magnetic radius ($R_m$), the
co-rotation radius ($R_{cr}$), and the light cylinder radius ($R_{lc}$). The
UV-optical luminous structures are the bright band at the inner edge of the
accretion disc and the surfaces of the Elvis structures on the irradiated
side of the dusty torus.}
\label{fig. 1}
\end{center}
\end{figure}

\newpage
\begin{figure}
\begin{center}
\plotone{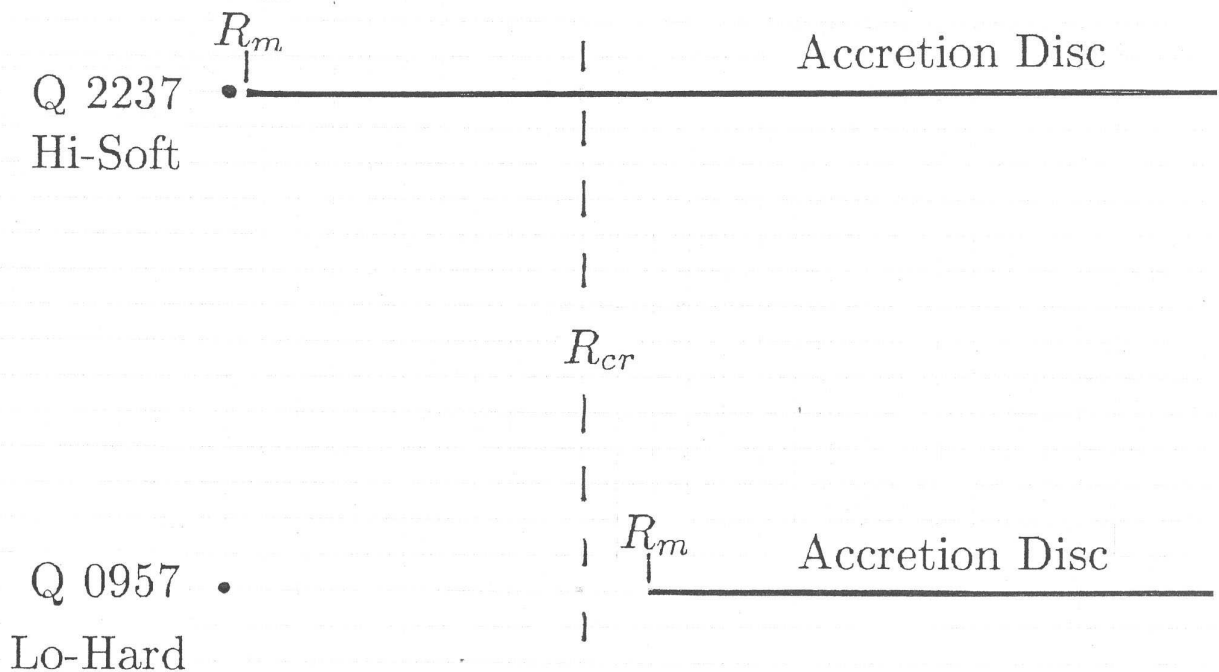}
\end{center}
\caption{A cartoon showing comparison of the inner structure of the two
quasars, emphasizing the important relationship of the magnetospheric
radius $R_m$ (at the inner edge of the accretion disc), to the co-rotation
radius $R_{cr}$, which is where the Keplerian orbital speed equals the
sweeping magnetic field speed. For this comparison the two quasars have
been rescaled to have the same co-rotation radius $R_{cr}$ to emphasize how
the accretion disc in Q2237 penetrates almost to the central MECO because
of its higher accretion rate, whereas
in Q0957 its luminous inner edge is outside of co-rotation. This difference
is the physical reason for the difference in spectral states between the
two quasars.}
\label{fig. 2}
\end{figure}

\end{document}